\documentclass[twocolumn]{elsart5p}

\usepackage{graphicx}
\usepackage{epsfig}
\usepackage{amssymb}
\usepackage{amsmath}
\usepackage{color}
\usepackage{xspace}
\usepackage{url}

\newcommand{\ccmskev}{\ensuremath{\text{counts}\,\text{cm}^{-2}\,\text{s}^{-1}\,\text{keV}^{-1}}\xspace}
\newcommand{\degree}{\ensuremath{^\circ}\xspace}

\begin{document}
\begin{frontmatter}

\title{Background study for the pn-CCD detector of CERN Axion Solar Telescope}

\author[Zara]{S. Cebri\'an}
\author[Zara]{A.~Rodr\'iguez\corauthref{cor}}
\corauth[cor]{Corresponding Author. Laboratorio de F\'isica Nuclear
y Altas Energ\'ias, Facultad de Ciencias, Pedro Cerbuna 12, 50009
Zaragoza, Spain. Phone number: 34 976761246. Fax number: 34
976761247} \ead{mara@unizar.es}
\author[TUD,MPE,GES]{M. Kuster}
\author[Zara]{B. Beltr\'an\thanksref{berta}}
\author[Zara]{H. G\'omez}
\author[HLL,pnsens]{R. Hartmann}
\author[Zara]{I. G. Irastorza}
\author[WHI]{R. Kotthaus}
\author[Zara]{G. Luz\'on}
\author[Zara]{J. Morales}
\author[Zara]{J. Ruz}
\author[HLL,MPE]{L.~Str\"uder}
\author[Zara]{J.~A. Villar}

\address[Zara]{Laboratorio de F\'{\i}sica Nuclear y Altas Energ\'{\i}as,
  Universidad de Zaragoza, 50009 Zaragoza, Spain}

\address[HLL]{MPI Halbleiterlabor, Otto-Hahn-Ring 6, 81739 M\"unchen,
  Germany}

\address[pnsens]{PNSensor GmbH, R\"omerstrasse 28, 80803 M\"unchen, Germany}

\address[WHI]{Max-Planck-Institut f\"ur Physik, F\"ohringer Ring 6, 80805
  M\"unchen, Germany}

\address[TUD]{Technische Universit\"at Darmstadt, IKP,
  Schlossgartenstrasse~9, 64289 Darmstadt, Germany}

\address[MPE]{Max-Planck-Institut f\"ur extraterrestrische Physik,
  Giessenbachstrasse, 85748 Garching, Germany}
\address[GES]{Gesellschaft f\"ur Schwerionenforschung, GSI-Darmstadt,
  Plasmaphysik, Planckstr. 1, D-64291 Darmstadt}

\thanks[berta]{Present address: Department of Physics, Queen's University, Kingston, Ontario K7L 3N6, Canada}

\date{}

\maketitle
\begin{abstract}
  The CERN Axion Solar Telescope (CAST) experiment searches for axions from
  the Sun converted into photons with energies up to around 10 keV via the
  inverse Primakoff effect in the high magnetic field of a superconducting
  Large Hadron Collider (LHC) prototype magnet. A backside illuminated pn-CCD detector in
  conjunction with an X-ray mirror optics is one of the three detectors used
  in CAST to register the expected photon signal. Since this signal is very
  rare and different background components (environmental gamma radiation,
  cosmic rays, intrinsic radioactive impurities in the set-up, \dots)
  entangle it, a detailed study of the detector background has been
  undertaken with the aim to understand and further reduce the background
  level of the detector. The analysis is based on measured data taken
  during the Phase~I of CAST and on Monte Carlo simulations of different
  background components. This study will show that the observed background level (at a rate of
  $(8.00\pm0.07)\times10^{-5}\,\ccmskev$ between 1 and 7 keV) seems to be
  dominated by the external gamma background due to usual activities at the
  experimental site, while radioactive impurities in the detector itself
  and cosmic neutrons could make just smaller contribution.
\end{abstract}


\end{frontmatter}

\section{Introduction}

Axions are a direct consequence of the Peccei-Quinn mechanism
\cite{peccei:77a} proposed to solve the so-called strong CP problem
(CP violation in strong interactions does not seem to exist in
nature, although the QCD Lagrangian contains CP-violating terms).
These particles could couple to two photons, which allows the
production of axions inside the hot plasma of stars via the
Primakoff effect. The expected flux of axions from the Sun has a
mean energy of $\sim$4.2 keV and virtually vanishes above 10 keV.
In the presence of a transverse magnetic field, solar axions could
be converted back to observable photons via the inverse Primakoff
effect, allowing axion detection on Earth \cite{sikivie:83a}.
\begin{figure*}
  \begin{center}
    \includegraphics[width=0.99\columnwidth]{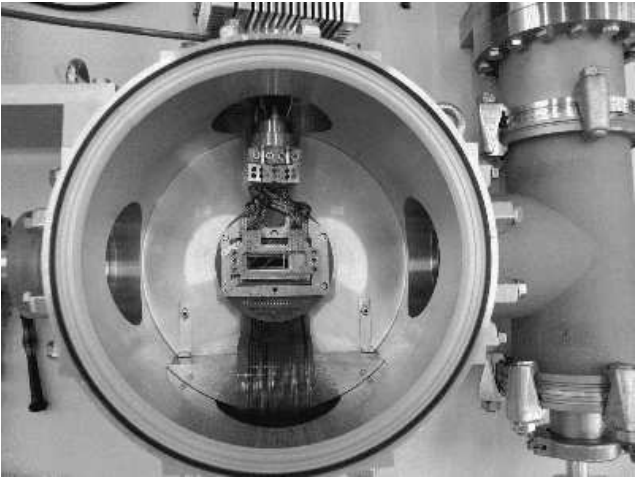}
    \hfill
    \includegraphics[width=0.96\columnwidth]{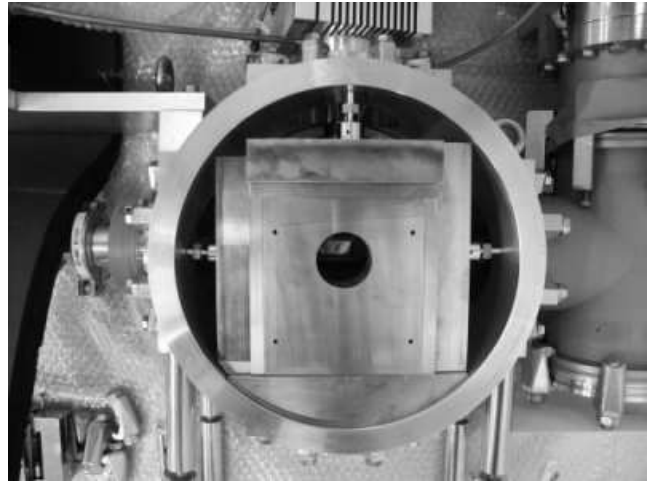}
\end{center}
  \caption{Left: The pn-CCD detector with its vacuum housing. The
    pn-CCD chip is the central black part surrounded by a gold
    plated cooling mask. Towards the top the connection to a cold finger of a
    Stirling cooler device is provided by copper leads. The connection to
    the detector electronics is provided by a flex-lead (towards the
    bottom). The vacuum vessel is made of Aluminum. Right: The same picture
    with the internal passive copper-lead shield being installed (view from
    the X-ray optics). The circular hole in the shield provides the
    aperture towards the X-ray optics. The external lead shield is not
    shown.}
  \label{fig:pnccd-dector}
\end{figure*}

The CERN Axion Solar Telescope (CAST) is intended to search for
solar axions based on this principle \cite{zioutas:99a,aalseth:06a}.
It uses a decommissioned $9.26\,\text{m}$ long LHC superconducting
magnet providing a $9\,\text{T}$ magnetic field, where the
probability of axion-to-photon conversion is proportional to the
length and magnetic field strength squared. Each bore of the twin
aperture magnet has a cross-sectional area of $14.5\,\text{cm}^{2}$.
The magnet is installed on a platform that permits horizontal
movement from azimuthal angle $46\degree$ to $133\degree$ and a
vertical movement of $\pm8\degree$. As a consequence, we can follow
the Sun three hours per day, 1.5 hours each during sunset and during
sunrise. The rest of the day is dedicated to background
measurements. Three X-ray detectors are installed at each end of the
magnet to search for excess X-rays coming from axion conversions
inside the magnet during alignment with the Sun. The detectors are a
Time Projection Chamber (TPC) \cite{autiero:06a}, a Micromesh
Gaseous Structure (MICROMEGAS) \cite{andriamonje:06a}, and a Charge
Coupled Device (CCD) in combination with a X-ray mirror telescope
\cite{xraytel:06a}. The CCD and MICROMEGAS detectors observe sunrise
axions, while the TPC detector, covering two magnet bores, looks for
sunset axions.
During 2003 and 2004 the experiment operated with vacuum inside the
magnet pipes (CAST Phase~I). Due to coherence effects this setup
allows us to explore an axion mass range up to $0.02\,\text{eV}$. No
signal above background was observed in 2003, implying an upper
limit to the axion-photon coupling of
g$_{a\gamma\gamma}\leq$1.16$\times 10^{-10}$ GeV$^{-1}$
\cite{zioutas:05a}. In order to extend the CAST sensitivity to
higher axion masses, the CAST experimental setup has been
transformed to be able to fill the axion conversion volume with a
buffer gas \cite{aalseth:06a}, probing the axion mass range up to
$0.8\,\text{eV}$ (CAST Phase~II). CAST started taking data in this
configuration in November 2005.

The CAST experiment is located at sea level at LHC point 8 of CERN
and the investigated axion signal is expected to be very rare.
Therefore, different background components will be merged with the
hypothetical signal in the detector data (environmental gamma
radiation, cosmic rays, intrinsic radioactive impurities in the
detector set-up, \dots). The goal of this work has been to study the
background of the pn-CCD detector of CAST with the aim to understand
its origin and to further reduce the background level to increase
the detector sensitivity.
Data acquired during the 2003 and 2004 periods
have been analyzed and a series of Monte Carlo simulations have been
performed using mainly the GEANT4 package \cite{geant4} to achieve
this goal. In particular, the response of this detector to photons
and neutrons has been studied and
a quantitative estimate of the contribution of different sources of
background (like external gamma and intrinsic radio-impurities in the
detector materials) to the overall background of the CCD detector has been
attempted.

The structure of this paper is the following: details of the CCD
detection system of CAST and its performance are presented in
Sec.~\ref{ccd} and the main sources of background expected in CAST
are described in Sec.~\ref{sources}, followed by a summary of the
CCD background properties in
Sec.~\ref{sec:measured-detector-background}. A description of the
simulations and results is shown in Sec.~\ref{simulations}.

\section{The CCD detector and the X-ray telescope} \label{ccd}
\label{sec:x-ray-telescope} The X-ray telescope of CAST is the most
sensitive of the three different detector systems currently in
operation, \cite{aalseth:06a}. It consists of a Wolter~I type X-ray
mirror optics \cite{wolter:52a,altmann:98a,egle:98a} focusing a
potential axion signal on a small area on a CCD detector which is
located in the focal plane of the optics. The CCD is a fully
depleted back side illuminated pn-CCD with a depletion depth of
$280\mu\text{m}$ and a pixel size of $150\times150\,\mu\text{m}^2$,
optimized for the $0.2$--$10\,\text{keV}$ energy range and currently
in use on board of ESA's XMM-Newton X-ray observatory
\cite{strueder:01a}.  The very thin SiO$_2$ radiation window of this
detector with a thickness of only $30\,\text{nm}$, results in a
quantum efficiency of $\approx 1$ between $1$ and $10\,\text{keV}$.
Such a device is operated for the first time in a low background
experiment. For an in-depth description of the design of the X-ray
telescope and its performance we refer to \cite{xraytel:06a}. The
long exposure time in CAST ($>3\,\text{years}$ of operation time)
provides a unique opportunity to study the long term background,
performance, and different shielding concepts for such a detector in
detail.

The combination of a detector with high quantum efficiency and a focusing
device already improves the signal to noise ratio of the experiment by a
factor of $\approx 12$ compared to a non-focusing system.
Further enhancement of the sensitivity could be achieved by adding a
passive shield consisting of a combination of internal (inside the
vacuum vessel) and external lead-copper components. As raw material
for the shield components we have chosen low activity oxygen free
copper and low activity lead, almost free of $^{210}$Pb. In
Fig.~\ref{fig:pnccd-dector} the CCD detector is shown as it is
integrated in its vacuum aluminum vessel with (right) and without
(left) internal shielding components.


\section{Background sources} \label{sources}
The CAST experiment is located at one of the buildings of the point
8 experimental area at CERN. Therefore, any component of the
environmental background at sea level can in principle affect the
measured background of the detectors of CAST.

The most important contribution is expected to be caused by external
gamma rays, produced mainly by primordial radio-nuclides like
$^{40}$K and the radioactive natural chains from $^{238}$U,
$^{235}$U and $^{232}$Th in laboratory soil, building materials and
experimental set-up as well as by $^{222}$Rn in air. Gamma rays of
cosmic origin make a negligible contribution. Activity levels of the
walls of the experimental hall where CAST operates were measured
using a germanium gamma spectrometer system \cite{dumont:04a} and
radon levels have been also monitored there during long periods.
Results from these measurements will be taken into account later on
to estimate the contribution of gamma rays to the total measured CCD
background counting rate.

Intrinsic radioactive impurities (either primordial or
cosmogenically induced) in the detector components and materials can
also make a relevant contribution in experiments looking for rare
event signals because of their alpha, beta and gamma emissions.
Since the CCD detector has an internal shielding made of lead-copper
within the vacuum chamber where it operates, impurities from the
external components can in principle be disregarded since they
should be greatly suppressed. Therefore, only impurities in the
materials composing the detector itself may be relevant and have
been taken into account in this study. Activities from the CCD
components were determined at the Canfranc Underground Laboratory
using an ultra-low background germanium spectrometer and will be
considered in evaluating their effect on the CCD measurements.

Cosmic rays on the Earth's surface are dominated by muons and
neutrons. While muon interactions (as those of other charged
particles) can be rejected with $\approx 100\%$ efficiency thanks to
their long ionizing tracks and the deposited energy, signals from
neutrons can contribute to the detector background. At sea level,
most of the neutrons come from cosmic rays. The neutrons induced by
cosmic muons or generated by fission or (alfa,n) reactions
\cite{heusser:95a} are in principle marginal. However, since
production of neutrons by muons is strongly enhanced in high Z
materials, muon-induced neutrons in the lead shielding have also
been analyzed.

In order to maximize the sensitivity of the CCD detector and
 to minimize the contribution of all background components,
  passive shields were installed and off-line rejection methods were
  applied. We want to point out that the signal-to-background ratio of the
  X-ray telescope is significantly better compared to the other two
  detectors in use in CAST due to the X-ray focusing system described in
  the previous section. Thus a sophisticated shielding concept including,
  e.g., a neutron shield seemed not to be of importance.

\begin{figure}
  \begin{center}
    \includegraphics[width=0.9\columnwidth]{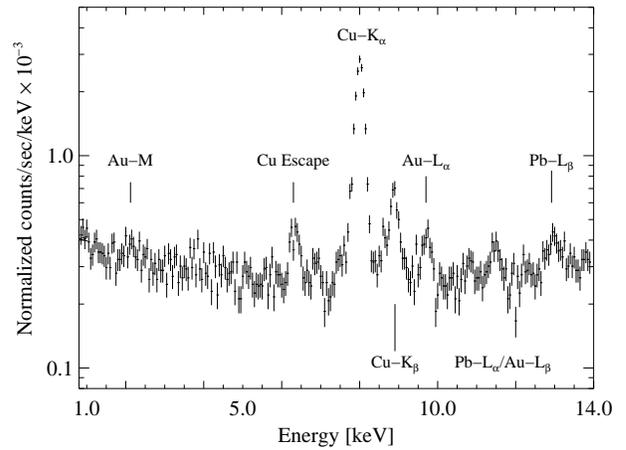}
  \end{center}
  \caption {A typical background spectrum measured with the CCD detector of
    CAST. Different fluorescent emission lines of material close to the CCD
    chip are apparent, e.g., the emission lines of copper, gold, and lead.}
  \label{measuredbackground}
\end{figure}

\section{Measured detector background}
\label{sec:measured-detector-background} During the last three years
of detector operation in CAST, we have acquired more than
$4298\,\text{h}$ of background data with the X-ray telescope which
have been partially analyzed. The 2003 background energy spectrum of
the CCD detector, shown above in Figure~\ref{measuredbackground},
exhibits a counting rate of $(11.5\pm0.2)\times10^{-5}\,\ccmskev$.
By adding internal and external shielding this level was reduced to
a raw counting rate of $(8.00\pm0.07)\times10^{-5}\,\ccmskev$.The
spectrum shows several characteristic emission lines of the
materials close to the CCD chip, e.g., the copper, gold, and lead
lines, on top of an almost flat continuum spectrum (Compton
scattered photons and secondary electrons \cite{xraytel:06a}). In
order to achieve a maximum sensitivity, the CCD detector is
optimized for the energy range from $1$ to $7\,\text{keV}$ which is
the range of interest for the axion search. The major goal of this
work is to explain the observed background in terms of different
contributions from environmental background, neutron or muon induced
background, and the contribution of natural radio-activity of the
detector components themselves.

Electronic noise of the detector and the so-called ``partial''
events have to be considered as intrinsic sources of background
\cite{popp:00a}. Since the electronic noise level of such CCDs is
far below $ \ll 100\,\text{eV}$, it is even for long exposure times
not important for our energy region of interest with a minimum
energy of $E_{\text{min}} \ge 0.7\,\text{keV}$. Calibrating spectrum
tell us that partial events contribute to it as a flat continuum
shown in Fig.~\ref{calibration}. These events arise from photons
absorbed in the detector near the Si-Si0$_2$ interface of the
radiation entrance window (see reference \cite{popp:00a}). For a
mono-energetic line such events contribute to the observed spectrum
with a constant fraction which is generally expressed in terms of a
peak-to-valley ratio (maximum of the photo-peak relative to the
level of the flat-shelf).  For the CAST pn-CCD detector this ratio
is of the order of $2000$. If we consider the Cu-K$\alpha$ line of
the background spectrum in Fig.~\ref{measuredbackground} we would
expect a contribution of at most $\approx
10^{-6}\,\text{counts}\,\text{s}^{-1}\,\text{keV}^{-1}$ to the
continuum spectrum between $1$--$7\,\text{keV}$ due to this effect.

\section{Background simulations} \label{simulations}
In order to estimate the contribution of external background and
  natural radioactivity to the overall background, a simulation tool for
  the CCD detector has been developed using the GEANT4 package.
 As a first step we cross-checked the reliability of the
  code by comparing calibration data like, e.g., measured quantum
  efficiency and calibration spectra with results we obtained from our
  simulations. The results will be presented in the following sections.
Simulations to understand the effect of the external environmental gamma
background and its attenuation in different shielding configurations are
presented and quantitative results of the estimate of the contribution of
this background to the CCD counting rates are shown.  Results regarding
neutrons, either from cosmic rays or muon-induced in lead, are also
collected.  Finally, an estimate of the contribution to the counting rate
of the CCD detector of the radioactive impurities of the detector materials
is made.

\subsection{The Code}
\begin{figure*}
  \begin{center}
    \includegraphics[width=0.55\textwidth]{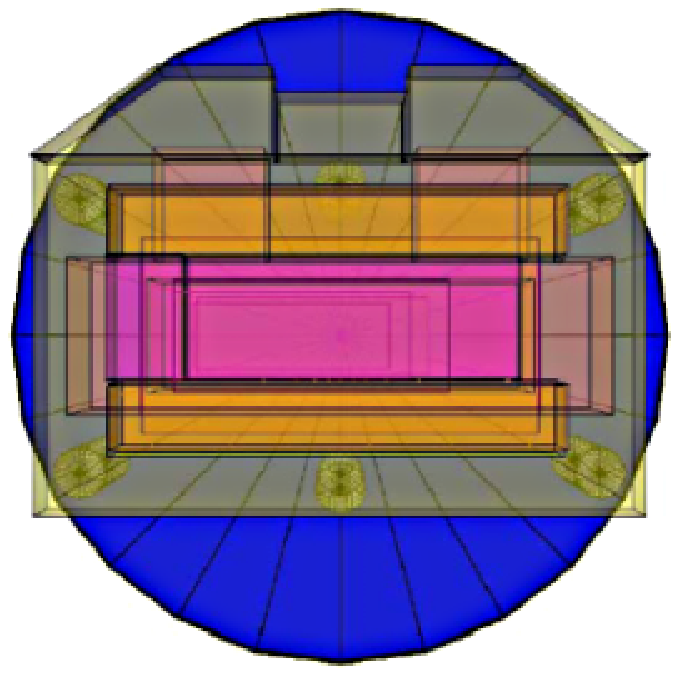}
    \hfill
    \begin{minipage}[b]{0.35\textwidth}
      \includegraphics[width=0.5\textwidth]{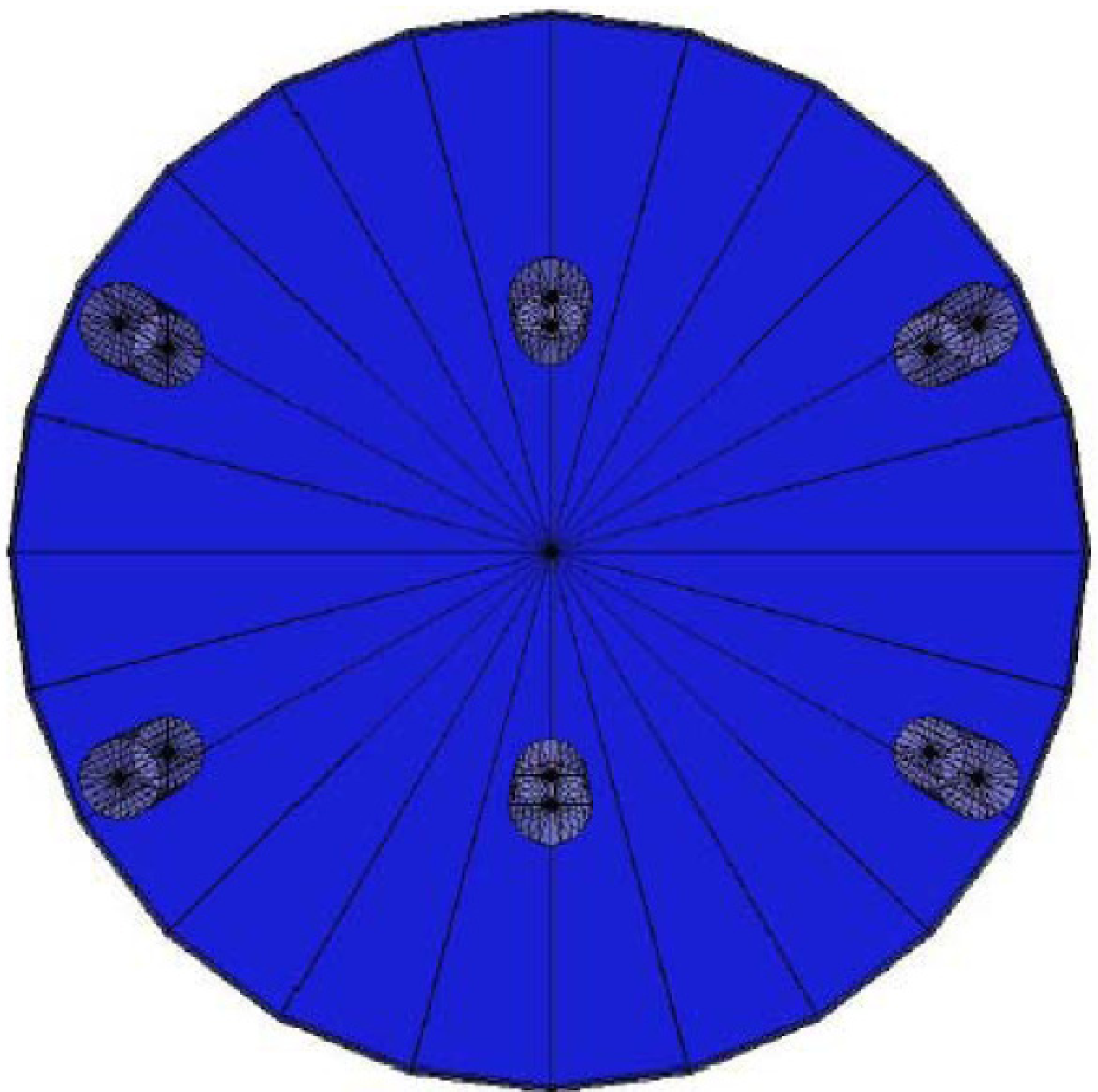}
      \includegraphics[width=0.5\textwidth]{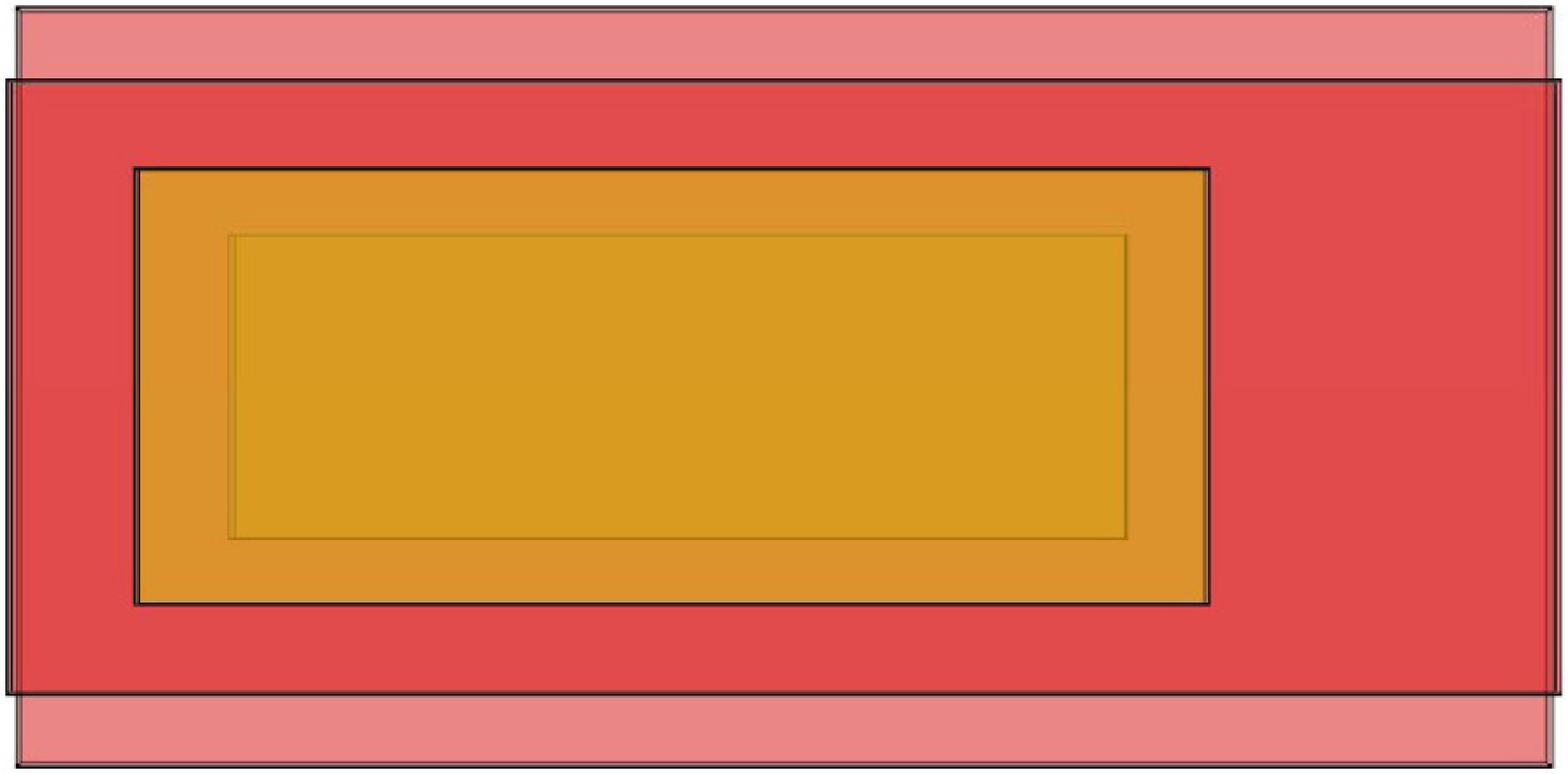}
      \includegraphics[width=0.5\textwidth]{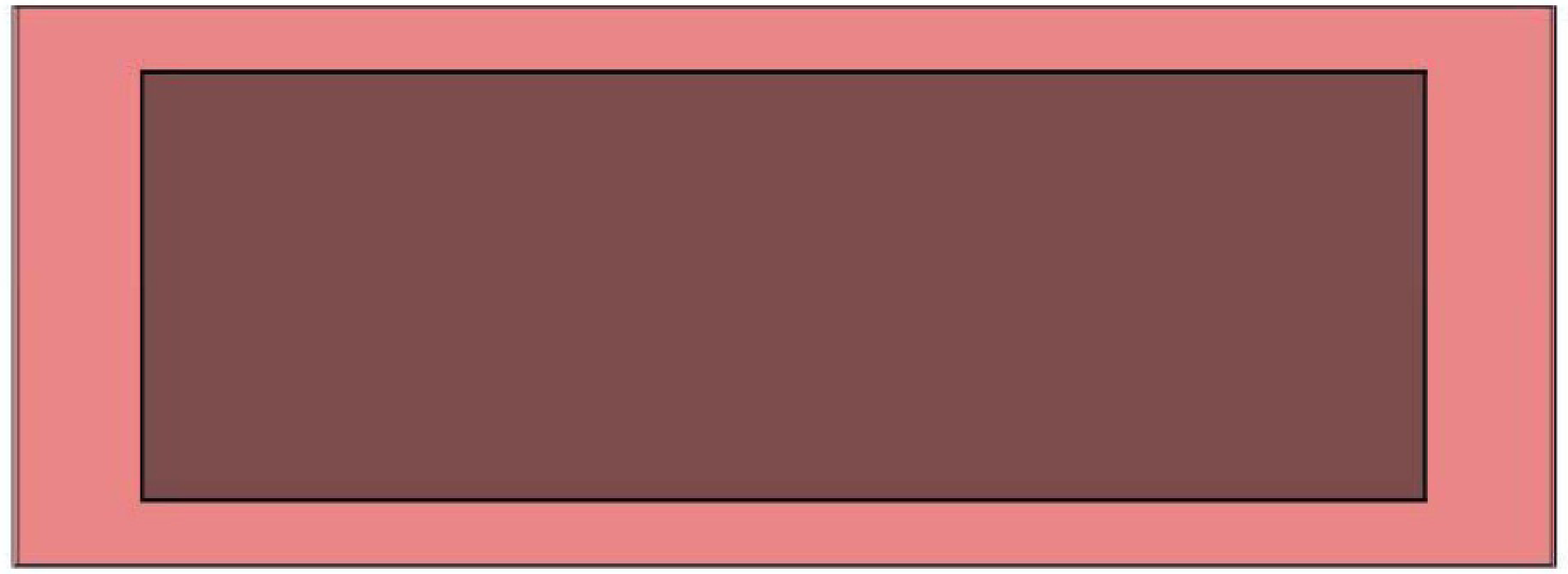}
      \includegraphics[width=0.5\textwidth]{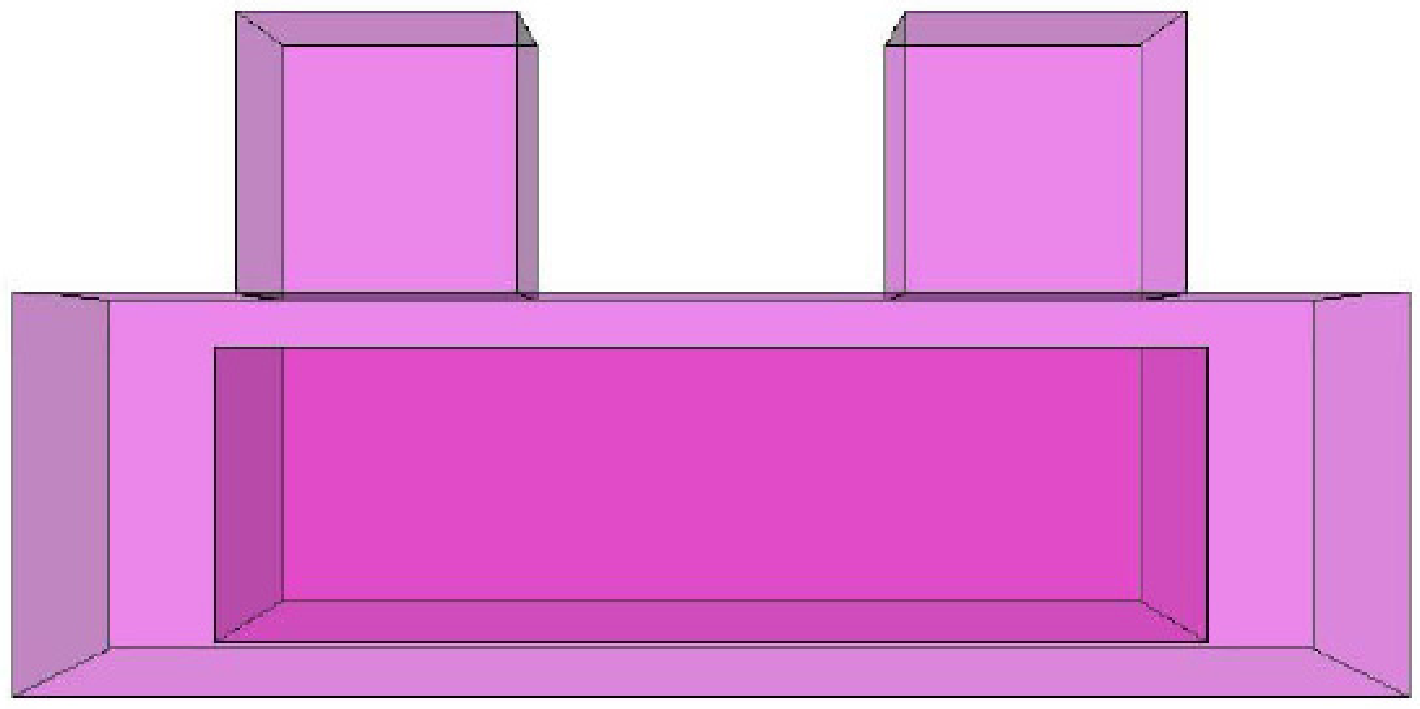}
      \includegraphics[width=0.5\textwidth]{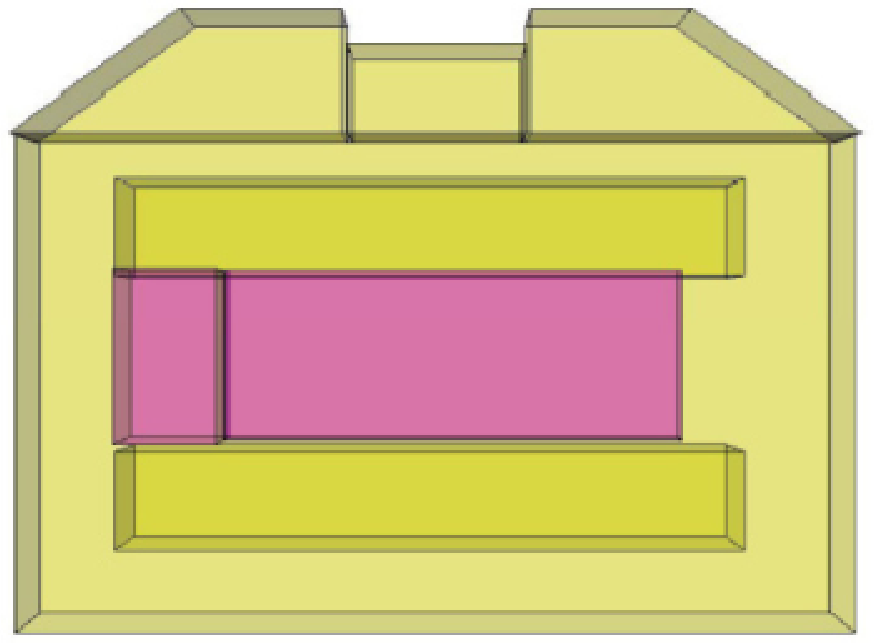}
    \end{minipage}
  \end{center}
  \caption{View of all simulated CCD detector components together
  (left)and independent views of each of them (right): electronic board, ceramics,
Si chip and front and back cooling mask.}
  \label{componentes}
\end{figure*}

\begin{figure*}
  \begin{center}
    \includegraphics[width=0.8\columnwidth]{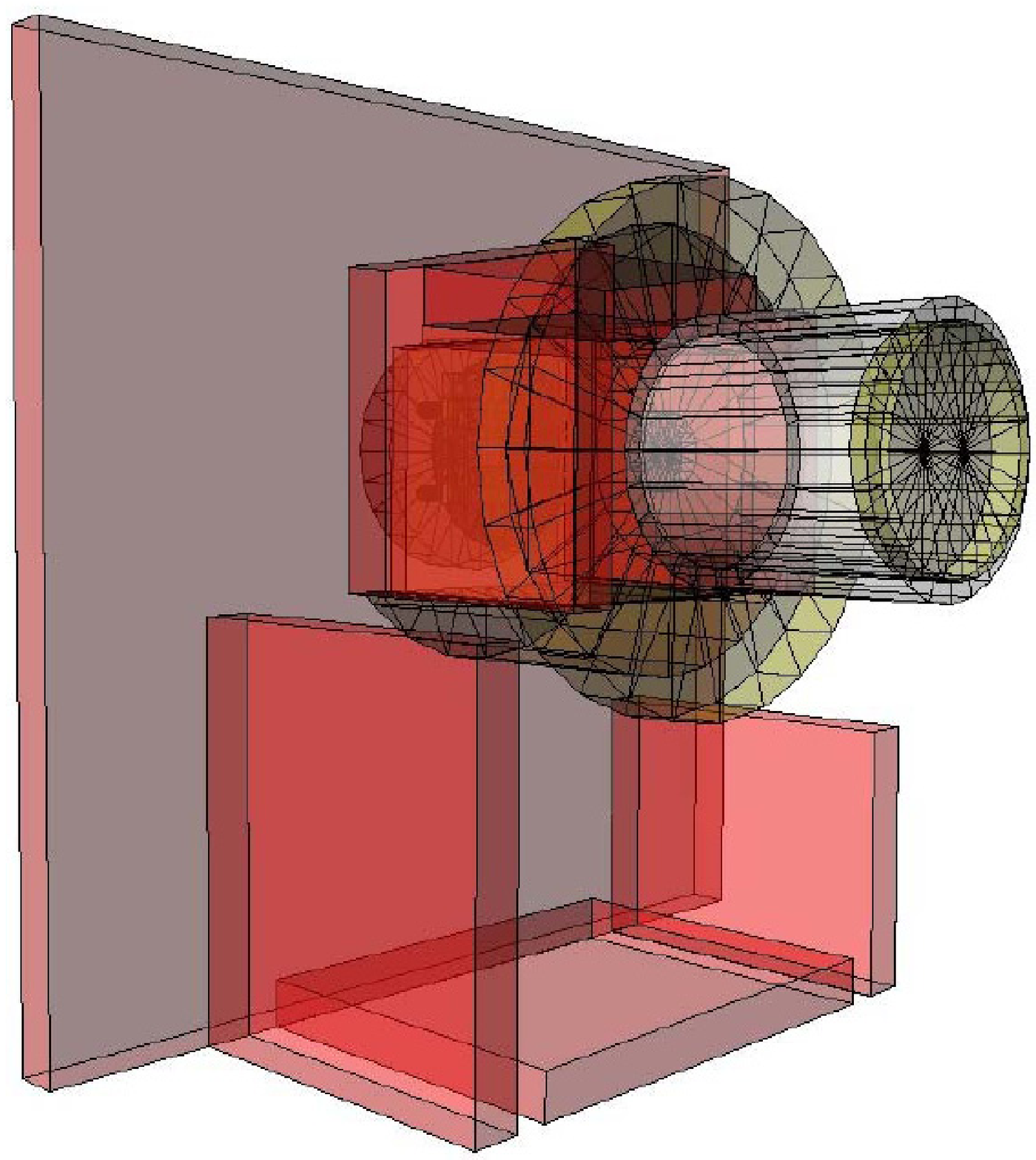}
    \includegraphics[width=0.8\columnwidth]{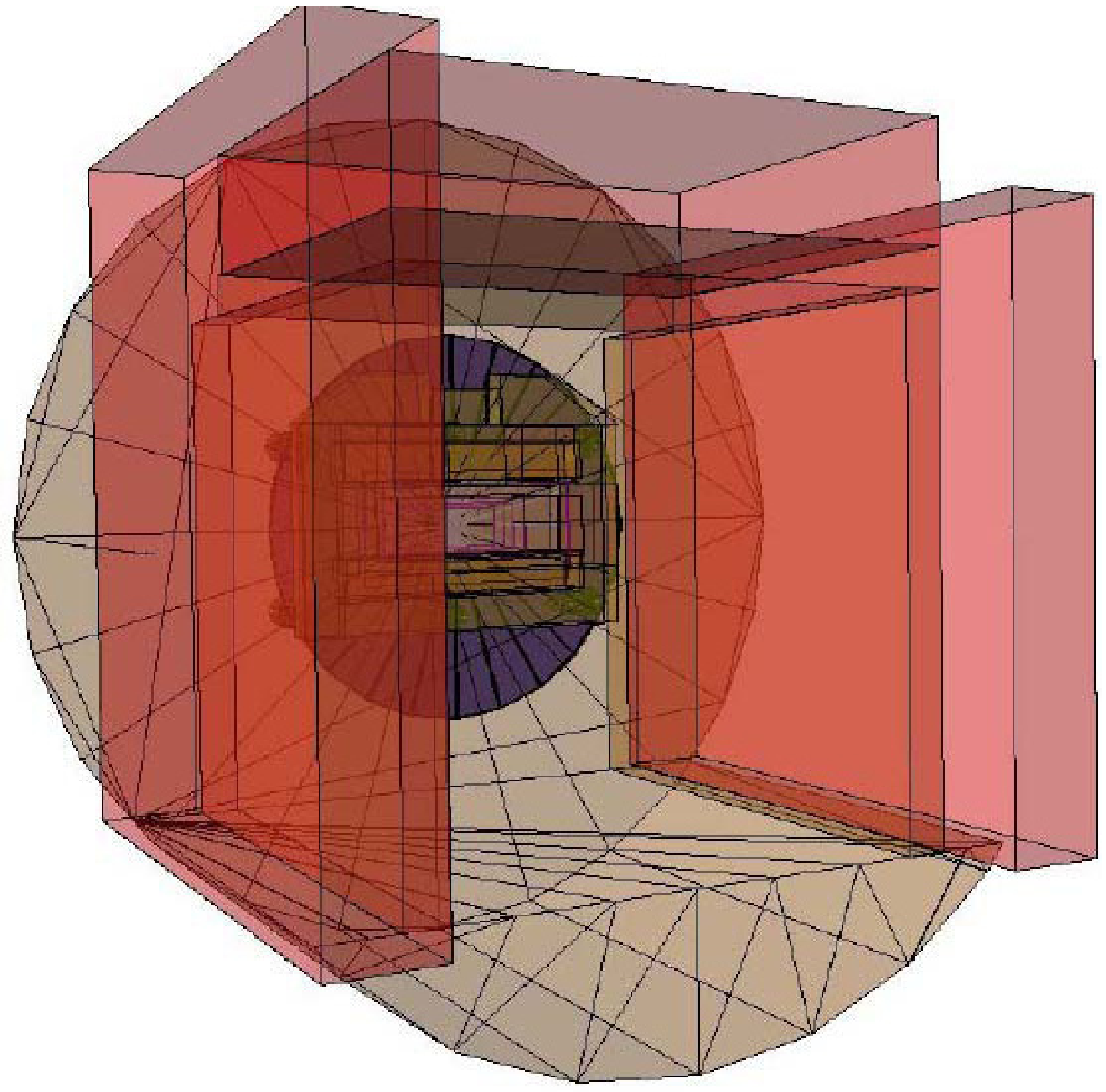}
  \end{center}
  \caption{Left: view of all simulated components, including shields,
    aluminum vessel and connection to telescope. Right: view of simulated
    components inside the aluminum vessel, including copper and lead
    shields and detector }
  \label{todo}
\end{figure*}

In the first simulations for neutrons and external gamma backgrounds
a simplified description of the detector was implemented just
considering the Si chip, the copper cooling mask and the printed
circuit board. A much more detailed geometry for the detector was
defined to carry out simulations of the radioactive impurities in
the detector components; shapes and sizes of Si chip (including
active area), ceramics, zero-force sockets, front and rear cooling
mask (including gold lining), and printed circuit board have been
reproduced as accurately as possible, keeping all the relations
between the sizes. Figure~\ref{componentes} shows views of the
simulated detector components. For both, the simplified and detailed
implementations of the detector, it is placed inside the aluminum
housing. Copper-lead shielding as well as the tube connecting to the
telescope have been considered. Figure~\ref{todo} shows the complete
simulated set-up.

Usual electromagnetic processes for photons and charged particles
have been taken into account using models specially developed for
low energy (valid above 250 eV), which is important to reproduce
fluorescence and emission of Auger electrons. Elastic and inelastic
scattering, capture and fission are considered for neutrons.
Radioactive decays are simulated also by GEANT4. Different
simulation cuts\footnote{Particles having an energy (or a path)
  lower than the fixed cut are no longer simulated and therefore they are
  directly absorbed.} have been defined for different kind of particles and
regions, to save computing time while having accurate results. For
each simulation, the spectrum of the energy absorbed in the CCD
detector (considering the real integration time of 72 ms) is
recorded.
\begin{figure}
  \includegraphics[width=0.97\columnwidth]{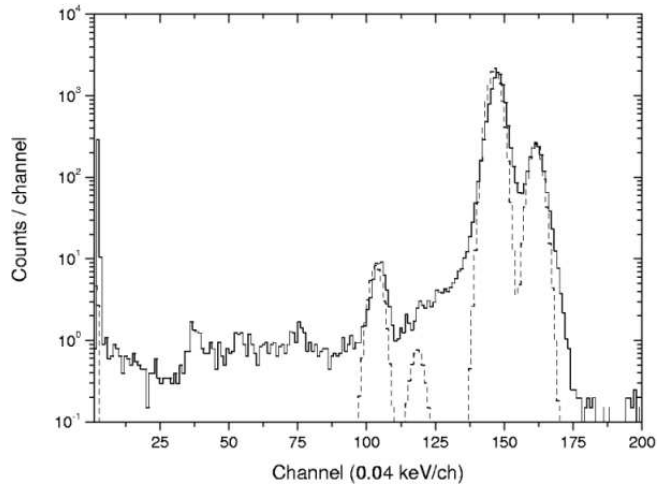}
  \caption {Measured (solid line) and simulated (dashed line)
    calibration spectra using a $^{55}$Fe source. The escape peaks
      corresponding to the Cu-K$\alpha$ and Cu-K$\beta$ photo peaks are
      well reproduced in energy and integral intensity. The continuum
      apparent up to channel 125 is not reproduced by the simulation (see
      text).}
  \label{calibration}
\end{figure}

\subsection{Verification Against Calibration Measurements}
An energy calibration with a $^{55}$Fe source was first simulated
and compared to the corresponding experimental spectrum.
Figure~\ref{calibration} shows this comparison. The area of the peak
at $\sim$4.2 keV due to the escape of Si X-rays produced following a
photoelectric interaction of 5.9 keV photons is quite well
reproduced (differences are of 1.5\%). The escape peak corresponding
to 6.5 keV photons is probably embedded in the background and the
other peaks. The continuum in the real calibration is much higher
than in the simulation. This discrepancy is related to the partial
charge collection of some events, which affects the response of the
CCD detector to low energy photons, as previously discussed in Sec.
\ref{sec:measured-detector-background} (details about this can be
found in Ref. \cite{popp:00a}). This effect accounts quantitatively
for the observed ratio between the continuum level and the area of
the biggest peak in calibration measurements, but its inclusion in
the simulation was disregarded since its influence on background
measurements can be neglected. Other effects which could produce
events in the continuum during calibrations (environmental
background, Compton continuum associated to peaks, energy losses of
photons within the source collimator or internal bremsstrahlung from
the $^{55}$Fe source) have been also studied, but their contribution
would be much smaller. In the measured spectrum a line from Al-K
X-rays at $\sim$1.5 keV is visible, due to the capsule where the
calibration source is placed. That capsule with aluminum filter
inside was not considered in the simulation.

\begin{figure}
    \includegraphics[width=0.9\columnwidth]{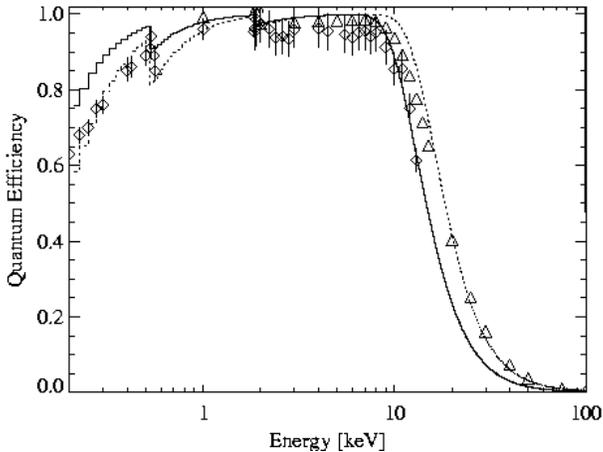}
  \caption {Simulated detection efficiency of the CCD to photons in the
    energy range from $0.2$--$100\,\text{keV}$ (dashed line and triangles). In comparison the measured
    quantum efficiency in the $1$--$15\,\text{keV}$ is shown as well (solid line and diamonds).}
  \label{intrinsicefficiency}
\end{figure}

In addition to $^{55}$Fe measurements, measurements to determine
detection efficiency as a function of energy (see
Ref.~\cite{strueder:01a}) were also simulated and results are
presented in Fig. \ref{intrinsicefficiency}. Independent simulations
have been carried out with slightly different geometry details.
Discrepancies between real data and simulations are of the order of
$\sim$4\% below 10 keV. Although no comparison with real data is
possible, high energy photons were also simulated obtaining for
instance an efficiency as low as $\sim$0.4\% for 100 keV photons.

\subsection{External Gamma Background}
\label{extgamma} In order to evaluate the contribution of the
environmental gamma background to the CCD counting rate and the
effect of the different shielding conditions, first the response of
the detector has been studied by means of photon simulation.

\subsubsection{Response}

\begin{figure}[tb]
    \includegraphics[width=0.47\textwidth]{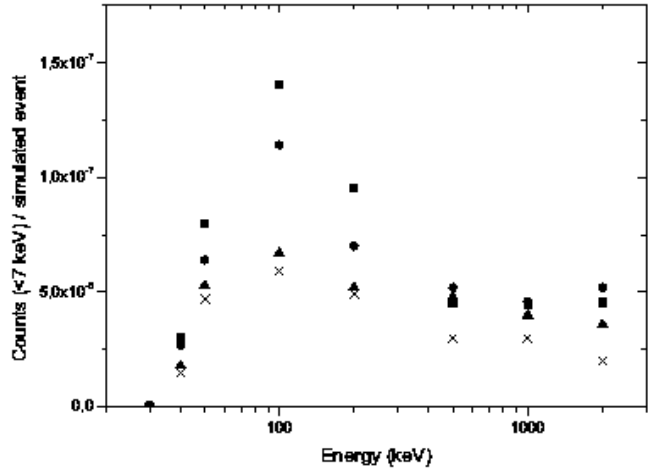}
  \caption {Simulated response of the CCD detector to the external gamma
    background: events in the region up to 7 keV per simulated photon as a
    function of the energy of the incident photons.  Four shielding
    conditions have been considered: with no shield (squares), with the
    internal copper shield (circles), with the internal copper+external
    lead shields (triangles) and with internal copper+external
    lead+internal lead shields (crosses).  }
  \label{gammaresponsesphere}
\end{figure}

The response of the CCD detector has been studied by simulating photons
with discrete energies from 30 up to 2000 keV from the surface of a sphere
containing all the components included in the simulation. Four different
geometries have been considered corresponding to the successive shieldings
added to the CCD set-up: no shield, internal copper shield, internal copper
shield and external lead shields and finally internal lead and copper
shields and external lead shield. Figure~\ref{gammaresponsesphere} presents
the ratio between the counts in the region up to 7 keV (to exclude X-rays
from copper) and the total simulated events for each one of the considered
initial energies of the incident photons. Results for each one of the four
different shielding conditions are plotted.

Gamma rays with energies around 100 keV contribute most to the CCD
background. However, the effect of this external gamma background
above 50 keV, according to Fig. \ref{gammaresponsesphere}, does not
seem to be very dependent on energy. This fact may be due to the
balance of two trends: low energy photons are less penetrating but
their contribution is higher while high energy photons can pass more
easily through the housing and the shielding but their contribution
is lower.

The average reduction achieved in the counting rate below 7 keV due to
environmental photons when using the complete shielding in the simulations
seems to be slightly underestimated compared to the reduction actually
observed in the CAST CCD detector (around a factor 3). Simplifications in
the implementation of the components of the set-up, uncertainties in the
response of the detector to high energy photons and/or non-isotropic
background sources as considered in the simulation could be responsible for
this discrepancy.

\subsubsection{Contribution to CCD counting rate}

Using this simulated response to external gamma including the effect
of the complete shielding (see Fig. \ref{gammaresponsesphere}), an
estimate of the contribution to the counting rate of the CCD
detector from the external gamma background has been attempted.
Results of the radon level measurements made by the CAST
Collaboration at the experimental site and the activity of the walls
of the experimental hall measured using a Ge spectrometer
\cite{dumont:04a} have been used as input in this estimate.

\begin{itemize}
\item{Radon contribution:} a mean radon concentration of $\sim$10 Bq m$^{-3}$ has
  been assumed, according to available measurements.  The rate of photons
  produced by $^{222}$Rn descendants from the air in successive spherical
  layers of radius R has been evaluated up to R=10 m. From Fig.
  \ref{gammaresponsesphere}, the number of counts up to 7 keV produced by a
  photon of fixed energy coming from a spherical surface is known; this
  information has been used to derive the contribution to counting rates of
  radon emissions of different energies from each of the spherical layers
  described, just properly rescaling the sphere used in the simulations to
  the corresponding air spheres. Total contribution from radon up to 7 keV
  turns out to be $\sim 10^{-6}\,\ccmskev$, almost negligible compared to
  the measured background rate in the CCD detector.
\item{Walls activity contribution:} specific activities for the North, South
  and East walls of the SR8 hall at CERN where the CAST experiment is
  running are known for $^{40}$K and for $^{232}$Th, $^{235}$U and
  $^{238}$U chains (in some cases, since equilibrium is broken for
  $^{238}$U chain two values are available up to and from $^{226}$Ra)
  \cite{dumont:04a}.
  The lower part of the East and South walls (around 1/4 of the total
  height) is made of composite wall while the upper part is made of concrete, having
  different specific activities according to measurements. The North wall
  contains partially metal. The rate of photons coming from the surface walls from
  each of the measured radioactive impurities has been evaluated using
  these specific activities, the wall dimensions and a mean density of 2.3
  g cm$^{-3}$. Uniform distribution of these impurities in the walls has
  been assumed and attenuation of photons through the walls has been taken
  into account. The contribution to the CCD counting rate of photons from
  each wall has been evaluated using again information from Fig.
  \ref{gammaresponsesphere}, rescaling the sphere used in the simulation to
  spheres having a radius given by the mean distance from the wall to the
  CCD detector.

  Table~\ref{gammacontribution} summarizes the counting rates obtained
  assuming the CCD detector at two different positions (height of 2.25 m,
  distance to South wall of 1.5 m and distance to East wall of 10 m for the
  first case and height of 3.70 m, distance to South wall of 0.5 m and
  distance to East wall of 9 m for the second one). Errors quoted in
  Table~\ref{gammacontribution} (around 6\%) come just from the error in
  the specific activities measurements. Considering the CCD detector at
  different positions along the tracking path when following the Sun does
  not change very significantly the contribution to the counting rates. A
  counting rate of around $3$--$4\times$10$^{-5}\,\ccmskev$ could be
  attributed to the measured activities of the experimental site walls.
  This could therefore be the bulk of the measured background rate.
\end{itemize}

\begin{table}
  \begin{center}
    \caption{Counting rates in units of $10^{-7}\,\ccmskev$ expected from natural
      radioactivity of the walls of the experiment area for two different
      positions.}
      \begin{tabular}{lrr}\hline
        Wall              &  Height $2.25\,\text{m}$  & Height $3.70\,\text{m}$ \\ \hline
        North             &  $22.27\pm0.48$ & $ 20.72\pm0.45$    \\
        East (lower part) &  $ 3.92\pm0.16$ & $  4.11\pm0.17$    \\
        East (upper part) &  $45.8\pm4.0$ & $ 52.9\pm4.6$      \\
        South (lower part)&  $21.96\pm0.92$ & $ 23.15\pm0.97$    \\
        South (upper part)&  $187\pm15$     & $317\pm26$         \\\hline
        Total             &  $281\pm16$     & $418\pm27$         \\\hline
    \end{tabular}
    \label{gammacontribution}
  \end{center}
\end{table}

\subsection{Neutrons}
\label{sec:neutrons} In a similar way as for photons, an evaluation
of the contribution of neutrons from different sources to the CCD
counting rate has been attempted, after studying the response of the
detector to neutrons using simulations.

\subsubsection{Response}
\label{sec:response} As for photons, the fraction of events detected
in the CCD detector in the region of interest has been evaluated
considering neutrons of different initial energies (from thermal up
to 10 MeV). The different capacity of nuclei and electrons to ionize
in the same medium compels us to take into account a relative
efficiency factor\footnote{It is defined as the ratio between the
amplitudes of pulses
  produced by nuclear and electronic recoils which deposit the same energy}
to compare energy spectra produced by neutrons and by photons. An
analytical expression based on the Lindhard theory and taken from Ref.
\cite{smith:90a} has been used for including this factor, which turns out
to be $\sim$0.28 for Si nuclear recoils of 10 keV and $\sim$0.36 for 100
keV. Nuclear recoils of up to $\sim$32 keV will be registered in the region
up to 10 keV of visible energy.

The main contribution of neutrons to the counting rates in the
energy region of interest comes from the elastic scattering off of
silicon nuclei. Simple kinematics says that for neutrons up to
$\sim$250 keV all the induced recoils deposit in the detector a
(visible) energy below 10 keV; neutrons with higher energies can
produce more energetic recoils.

\begin{figure}[tb]
    \includegraphics[width=0.35\textwidth]{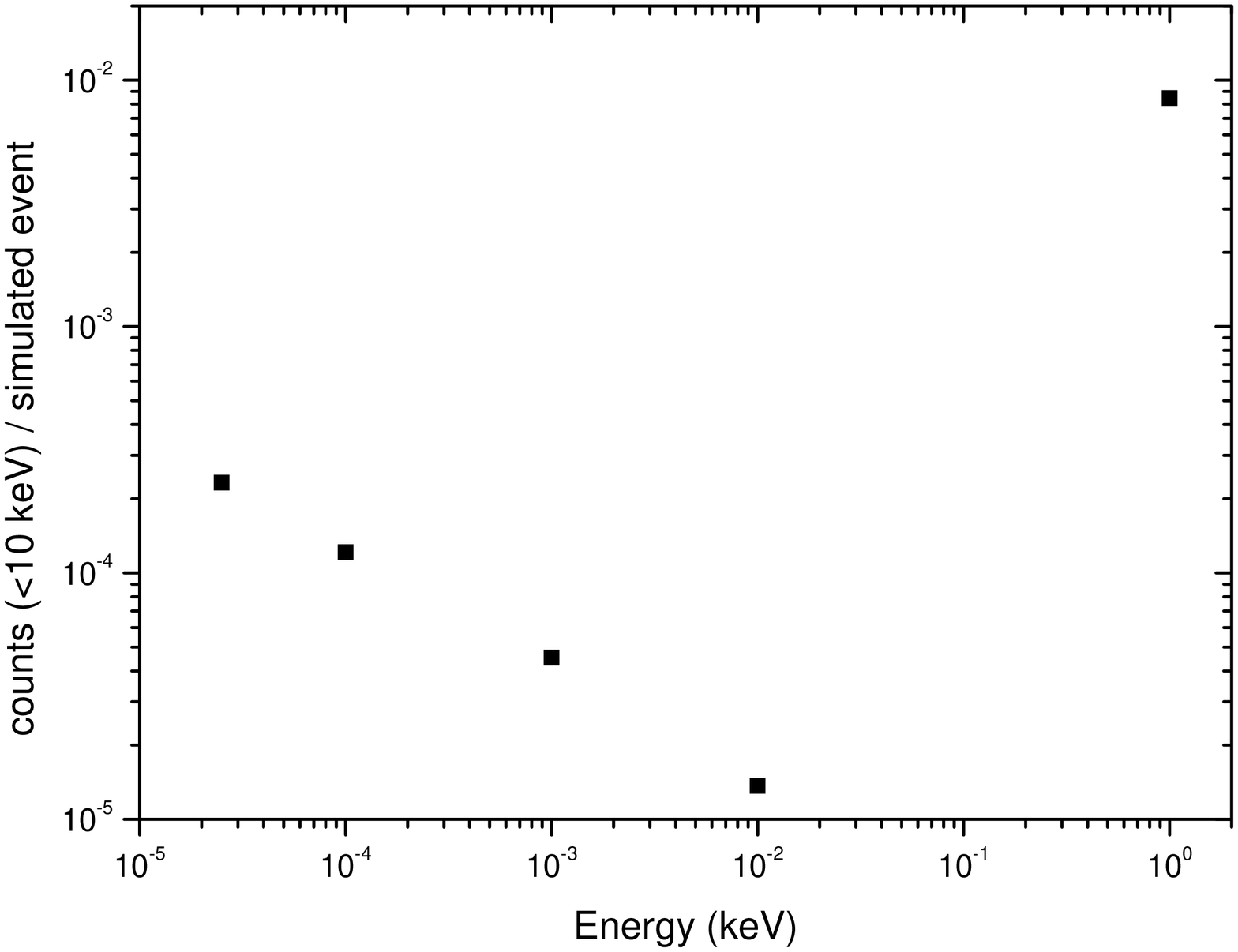}
    \includegraphics[width=0.35\textwidth]{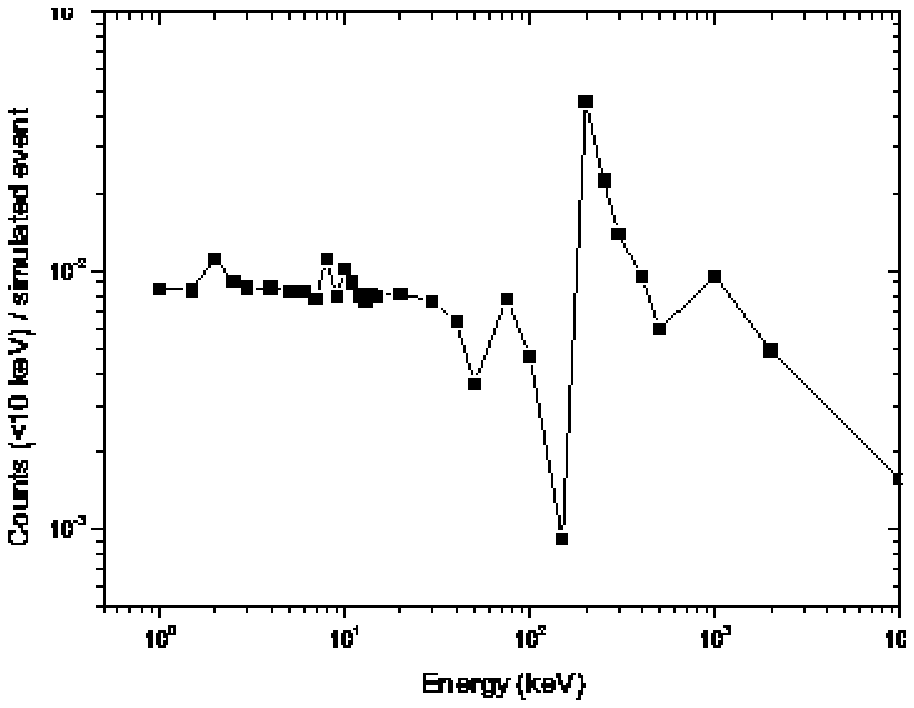}
  \caption {Simulated response of the CCD to neutrons in the low energy
    region: events registered up to 10 keV (visible) per simulated neutron
    as a function of the energy of the incident neutrons. On top, for
    neutrons up to 1 keV, at the bottom for neutrons above 1 keV.}
  \label{neutronresponse}
\end{figure}

Plots in Fig.~\ref{neutronresponse} present the ratio between the
events recorded in the region up to 10 keV of visible energy and the
total simulated events, as a function of the energy of incident
neutrons (up to 1 keV on top and above 1 keV at the bottom). The
maxima appearing in Fig. \ref{neutronresponse} (bottom) are due to
resonances in the cross-sections for neutron elastic scattering on
silicon. Since the cosmic-ray neutron energy spectrum is peaked at
thermal energies ($\sim$0.025 eV) \cite{hess:59a}, the response of
the CCD detector to thermal neutrons has been also evaluated. The
contribution of thermal neutrons to the background in the energy
region of interest (see Fig. \ref{neutronresponse}, top) is at least
one order of magnitude lower than the contribution of neutrons above
1 keV. Events induced by these very low energy neutrons are due not
to nuclear recoils following elastic scattering but to radiative
capture around and/or inside the detector; consequently, any gamma
shield will help to further reduce the contribution of thermal
neutrons. Below 1 keV, the contribution to background increases when
energy decreases because of the higher relevance of neutron capture.

\subsubsection{Contribution to CCD counting rate}
\label{sec:contribution-to-ccd-counting-rate} A rough estimate of
the contribution of the environmental neutron background to the CCD
counting rate in the region of interest has been attempted
considering the neutron fluxes reported in Table~\ref{nflux}
(corresponding to integrals in different regions of energy of the
neutron flux at sea level according to Refs.
\cite{heusser:95a,hess:59a,ziegler:98a}). Assuming for each one of
the four energy intervals in Table~\ref{nflux} an average value for
the fraction of events detected in the region of interest (from the
response plotted in Fig. \ref{neutronresponse}), the counting rate
due to the external neutron background in the region up to 10 keV
has been estimated to be $6\times10^{-6}\,\ccmskev$. Since this
number is one order of magnitude lower than the experimental
background levels of the CCD detector, it seems that environmental
neutrons are not very significant at the present level of
sensitivity.

\begin{table}
  \caption {Measured neutron fluxes in units of
    $10^{-3}\,\text{cm}^{-2}\,\text{s}^{-1}$ at sea level for
    different energy ranges.\label{nflux}}
  \begin{center}
    \begin{tabular}{rcc}\hline
      Energy Range           & Neutron Flux  & Reference      \\\hline
      $<1\,\text{eV}$        & $\sim 1$             & \cite{hess:59a}    \\
      $0.4$--$0.1\,\text{MeV}$& $2.9$               & \cite{heusser:95a} \\
      $0.1$--$1\,\text{MeV}$ & $1.6$                & \cite{heusser:95a} \\
      $1$--$10\,\text{MeV}$  & $1.7$                & \cite{heusser:95a} \\
      $>10\,\text{MeV}$      & $5.6$                & \cite{ziegler:98a} \\ \hline
    \end{tabular}
  \end{center}
\end{table}
At sea level, neutron production by capture of negative muons is
strongly enhanced especially in high Z materials. Therefore, an
estimate has been made also for the muon-induced neutrons in the
lead shielding of the CCD detector. Induced neutrons come mainly
from evaporation and therefore their energy spectrum is peaked
around 1 MeV and very reduced above 5 MeV. Using a measured total
muon flux of $\sim50.3\times10^{-3}\,\text{cm}^{-2}\,\text{s}^{-1}$,
FLUKA simulations \cite{fluka:01}\cite{fluka:02} give a yield for
neutrons of around
$\sim1\times10^{-3}\,\text{cm}^{-2}\,\text{s}^{-1}$ which is more
than one order of magnitude lower than the flux due to environmental
neutrons.Therefore, the installation of the present lead shield is
not a problem from the point of view of muon-induced neutrons.

It is worth noting that the values quoted for the estimated
contribution to the CCD counting rate of the external gamma and
neutron backgrounds must be considered as just an indication of the
order of magnitude of these contributions. Both particle fluxes and
simulated detector responses (the key ingredients of these
estimates) have important uncertainties which in many cases are
difficult to be quantified. Some facts that contribute to these
uncertainties are that high energy photon and neutron simulations
have not been checked against experimental data, that radon levels
are known to have important fluctuations in time or that the neutron
flux at sea level depends on different factors not taken into
account.  Nevertheless, results have been useful to identify the
main background sources of the CCD detector.

\subsection{Intrinsic radioactive impurities}
\label{impurities}
An evaluation of the contribution of the internal radioactive impurities in
the components of the CCD to the counting rate measured in this detector
has been made based on some measurements of radiopurity levels of several
components and simulation of the effect of these activities. These
estimates can help in the development and construction of a new CCD
detector optimized from the radioactive point of view.

The levels of radioactive impurities in the main components of a CCD
detector equivalent to that actually used in the CAST experiment
(circuit board, sockets, ceramics, Si detector and cooling mask)
were measured in the Canfranc Underground Laboratory in Spain using
a special ultra-low background germanium detector and can be found
in a database of radiopurity of materials\footnote{Available at
\url{http://radiopurity.in2p3.fr/}}inside the ILIAS program
(Integrated Large Infraestructures for Atroparticle Science).
Activities come mainly from the radioactive chains $^{235}$U,
$^{238}$U, $^{232}$Th and the isotope $^{40}$K. For the Si chip and
the cooling mask pieces only upper limits could be derived for the
impurities of the radioactive chains $^{235}$U, $^{238}$U and
$^{232}$Th. For the chip, also for $^{40}$K only a limit was
deduced. $^{238}$U chain is broken at the level of $^{226}$Ra for
the sockets and ceramic pieces.

\subsubsection{Simulations}
\begin{figure}
    \includegraphics[width=0.35\textwidth]{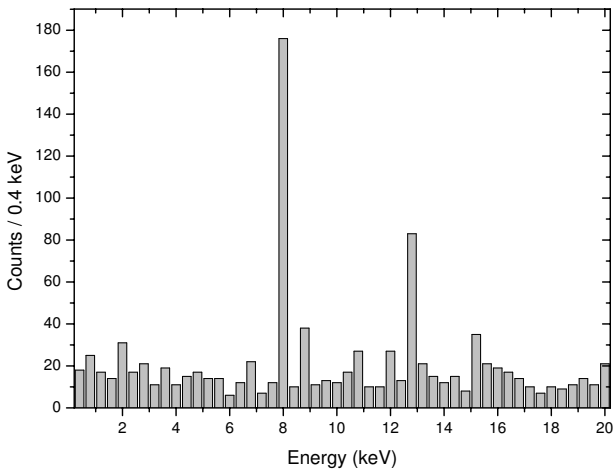}
  \caption{Simulated energy spectrum registered in the CCD for
    $^{232}$Th impurities in the ceramics.}
  \label{spcceramics}
\end{figure}

For each component of the CCD detector, complete decays of the
radioactive chains $^{235}$U, $^{238}$U and $^{232}$Th, assumed in
secular equilibrium, and the isotope $^{40}$K have been simulated
considering impurities uniformly distributed in the materials. As a
result, a collection of energy spectra corresponding to each
impurity in each component has been obtained. Just as an example,
Fig. \ref{spcceramics} shows the simulated spectrum registered in
the CCD detector for $^{232}$Th impurities in the ceramic pieces (it
must be noted that no normalization has been made in this plot).

Peak structures observed in the real CCD spectrum (see Fig.
\ref{measuredbackground}) at energies slightly higher than those of the
region of interest are partly reproduced by these simulations:

\begin{itemize}
\item In most of the simulated spectra for the different components, the
  X-ray K$_{\alpha}$ line from copper at 8.0 keV is clearly imposed on a
  flat background. It is due to photoelectric interactions of more
  energetic photons in copper pieces. A much smaller line at 8.9 keV is
  visible too due to much less intense copper K$_{\beta}$ X-rays.
\item X-rays from gold (L$_{\alpha}$ at 9.7 keV and L$_{\beta}$ at 11.5
  keV) are also present since the gold lining of the cooling mask pieces
  where they are generated has been properly included in the simulations.
\item Other low energy gamma lines and especially X-rays between 10 and 15
  keV emitted by daughter nuclei in the radioactive chains have been
  identified in the simulated spectra for the components closest to the Si
  chip. But the X-rays observed in the measured CCD spectrum and
  presumably attributed to ordinary solders in the electronic board
  containing $^{210}$Pb are not reproduced since these solders have not
  been taken into account in the simulations.
\item In the simulated spectrum for $^{40}$K in the chip, a peak at $\sim$3
  keV is seen due to the X-rays following the electronic capture in this
  isotope. This peak is not observed in the real spectrum because it is
  probably hidden by the continuum background, which points to a completely
  negligible $^{40}$K impurity in the Si chip.
\end{itemize}

\subsubsection{Results}
\label{sec:results} Table~\ref{totals} shows the total contributions
to the CCD counting rate of each simulated component between 1 and 7
keV. Contribution from the Si chip and the cooling mask pieces must
be considered just an upper limit, since only bounds for activity
were set for these components, as stated before. Although the
screening of radiopurity was made only for the front part of the
cooling mask, equal activity has been considered for the rear part
assuming that the same kind of copper was used for both pieces.
Contribution from $^{238}$U in the sockets and ceramic pieces is
overestimated since the assumed secular equilibrium in the chain is
not true, having from $^{226}$Ra onwards a lower activity. The
errors quoted in Table~\ref{totals} (between 5 and 10\%
approximately) include the statistical error associated with the
simulation and also the error of the measured activity of the
pieces. In general, errors from the activity dominate, although
statistical errors are not always negligible because of the large
execution time of simulations.

\begin{table}
  \caption{Total contribution to CCD counting rate between 1 and 7
keV from natural radioactivity of the individual detector components
in units of
    \ccmskev. We considered the $^{238}$U, $^{235}$U, $^{232}$Th, and
    $^{40}$K decay chains.}
  \label{totals}
  \begin{center}
    \begin{tabular}{lr} \hline
      Detector Component & \multicolumn{1}{c}{Differential Flux}    \\\hline
      CCD Board          &   $(5.83\pm0.41)\times10^{ -7}$          \\
      CCD Chip           & $<2.20\times10^{ -5}$          \\
      Ceramics           & $(1.17\pm0.14)\times10^{ -6}$            \\
      Sockets            & $(1.34\pm0.15)\times10^{ -6}$            \\
      Front Cooling Mask & $<3.60\times10^{ -7}$           \\
      Rear Cooling Mask  & $<2.61\times10^{-7}$            \\\hline
    \end{tabular}
  \end{center}
\end{table}
These estimates indicate that the components whose radioactive
impurities have been actually measured (CCD board, ceramics and
sockets) produce $(3.05\pm0.20)\times10^{-6}\,\ccmskev$ in the
region up to $7\,\text{keV}$. Considering the upper limits of
radio-impurities for all the other components, the total
contribution to background would be $2.6\times10^{-5}\,\ccmskev$.
Compared to the measured background level in the CCD detector
between $1$ and $7\,\text{keV}$,
$(8.00\pm0.07)\times10^{-5}\,\ccmskev$, impurities from CCD board,
ceramics, and sockets account for $\sim$4\% of the observed counting
rate; including all the other simulated impurities, up to $\sim$33\%
of the counting rate would be justified at most.

\section{Conclusions}
\label{sec:conclusions} The CCD detector of the CAST experiment
looks for the very rare signal of solar axions, consisting of
photons peaking around 4 keV. Different background components
entangle the expected signal.  A Monte-Carlo simulation tool for
this detector has been developed and used for the most relevant
background sources with the aim to help in the understanding of the
origin of observed events and constructing a plausible background
model for the measured counting rate
$(8.00\pm0.07)\times10^{-5}\,\ccmskev$ between $1$ and
$7\,\text{keV}$.

Using the simulated response of the CCD detector to the external gamma
background (including the effect of shieldings), an estimate of the
contribution of this background component has been attempted, finding that
measured radon levels in the air of the CAST site could produce on average
$\sim$1\% of the registered counting rate, while however just the measured
activities from $^{40}$K and $^{232}$Th and $^{238}$U chains in the walls
of the CAST hall could justify more than 50\% of this counting rate.

The response of the CCD to neutrons of different energies has also
been simulated. Based on this response and typical fluxes of
neutrons at sea level, a rough estimate of the contribution to the
CCD counting rate of environmental neutrons has been made. At the
present level of sensitivity, these neutrons do not seem to be a
very relevant source of the CCD background, producing just a few per
cent of the observed counting rate. Contribution from muon-induced
neutrons in the present lead shieldings has been checked to be
negligible.

Finally, the contribution to the CCD counting rate of the internal
radioactive impurities of the main detector components has been
simulated using the activities measured at the Canfranc Underground
Laboratory with an
 ultra-low background germanium detector. This
contribution could justify at most 33\% of the measured counting
rates.

Taking into account all these results, a quite complete model for the
background measured by the CCD detector has been obtained. Other possible
relevant sources of background, not evaluated up to now, are thought to be
the radioactive impurities from the soil and from massive components of the
experimental set-up.

\section{Acknowledgments}
\label{sec:acknowledgment} This work has been performed in the CAST
collaboration. We thank our colleagues for their support.  Furthermore, the
authors acknowledge the helpful discussions within the network on direct
dark matter detection of the ILIAS integrating activity (Contract number:
EU-RII3-CT-2003-506222). Our gratitude also to the group of the Canfranc
Underground Laboratory for material radiopurity measurements. This project
was also supported by the Bundesministerium f\"ur Bildung und Forschung
(BMBF) under the grant number 05 CC2EEA/9 and 05 CC1RD1/0, by the
Virtuelles Institut f\"ur Dunkle Materie und Neutrinos -- VIDMAN, and by
the Spanish Ministry of Education and Science under contract FPA2004-00973.

\bibliography{mnemonic,cast,conferences,detback,detector,darkm}

\bibliographystyle{elsart-num}


















\end{document}